\documentclass[aps,twocolumn,superscriptaddress,showpacs]{revtex4}
\usepackage[dvips]{graphicx}
\usepackage[]{caption}
\begin{document}

\title{Phase Behavior of Columnar DNA Assemblies}

\author{H.~M. Harreis}
\affiliation{Institut f{\"u}r Theoretische Physik II, 
Heinrich-Heine-Universit\"at D\"usseldorf,
Universit\"atsstra{\ss}e 1, D-40225 D\"usseldorf, Germany}
\author{A.~A. Kornyshev}
\affiliation{Institut f{\"u}r Theoretische Physik II,
Heinrich-Heine-Universit\"at D\"usseldorf,
Universit\"atsstra{\ss}e 1, D-40225 D\"usseldorf, Germany}
\affiliation{Institute for Materials and Processes in Energy Systems III,
Research Centre J{\"u}lich, 
D-52425 J{\"u}lich, Germany}
\author{C.~N. Likos}
\affiliation{Institut f{\"u}r Theoretische Physik II, 
Heinrich-Heine-Universit\"at D\"usseldorf,
Universit\"atsstra{\ss}e 1, D-40225 D\"usseldorf, Germany}
\author{H. L{\"o}wen} 
\affiliation{Institut f{\"u}r Theoretische Physik II, 
Heinrich-Heine-Universit\"at D\"usseldorf,
Universit\"atsstra{\ss}e 1, D-40225 D\"usseldorf, Germany}
\author{G. Sutmann}
\affiliation{John von Neumann
Institute for Computing,
Research Centre J{\"u}lich,
D-52425 J{\"u}lich, Germany}

\date{\today, submitted to \sl{Physical Review Letters}}

\pacs{82.35.Rs, 64.70.-p, 87.14.Gg, 82.70.Dd}

\begin{abstract}

The pair interaction between two stiff parallel linear 
DNA molecules depends not only on the distance
between their axes but on their azimuthal orientation. 
The positional and orientational order in columnar
B-DNA assemblies in solution is investigated, based on
the DNA-DNA electrostatic pair potential
that takes into account DNA helical symmetry 
and the amount and distribution of adsorbed
counterions. 
A phase diagram obtained by lattice sum calculations predicts a variety of
positionally and azimuthally 
ordered phases and bundling transitions strongly depending on the 
counterion adsorption
patterns.
\end{abstract}

\maketitle

DNA is a polyelectrolyte molecule. 
In aqueous electrolyte solutions, cations along its helices 
dissociate from it and dissolve into the mixture, leaving behind
negative charges that reside on the phosphates of the DNA backbone. 
According to the Manning condensation theory, a fraction
of the cations condenses into the
Bjerrum layer near the molecular surface \cite{manning:biophys:78}. 
If some of the ions specifically
adsorb onto DNA, its surface could be almost fully neutralized \cite{wison:79}
or even overcharged \cite{pelta:96}.
Far from its axis, DNA can be apprehended as a
charged cylinder. If the charge were continuously smeared, 
there would be only an electrostatic
repulsion between two molecules, exponentially screened by the electrolyte. However, the net
distribution of charge on the molecules is not homogeneous 
and this can dramatically alter the
interaction potential at intermediate distances. 
Indeed, in order to condense DNA in an aggregate, 
one has either to apply osmotic stress \cite{rau:etal:pnas:84}
or use condensing agents, 
such as salts with Mn$^{\rm{2+}}$, Cd$^{\rm{2+}}$,
spermidin, protamine or cobalt hexammine \cite{bloomflield:96} cations. 
These cations are known
to specifically adsorb on DNA, predominantly 
in the DNA {\it grooves} \cite{tamir:etal:93}. Other
counterions, such as, e.g., Ca$^{\rm{2+}}$ or Mg$^{\rm{2+}}$, 
that have strong affinity to
phosphates and adsorb preferentially on the {\it strands} 
do not induce DNA aggregation. Obviously,
one effect of these specifically adsorbing counterions 
is the reduction of the net charge on the
DNA. However, were this to be the only effect, 
it would have been hard to explain the observed 
sensitivity to the sort of counterions of DNA condensation \cite{bloomflield:96}
and of the
mesomorphism of resulting aggregates \cite{podgornik:cocis:84}.

Recently a new explanation of the features of 
DNA aggregation was suggested \cite{kl:prl:99}
resting on a Debye-Bjerrum theory of electrostatic
interaction between helical macromolecules \cite{kl:JCP:97}.
The theory offered first a
formalism for a description of  
interaction between cylindrical molecules (with parallel axes) for
arbitrary surface charge distributions on the molecules \cite{kl:JCP:97}. 
Then it explored its
consequences for helical charge distributions, including those 
typical for double stranded B- and
A-forms of DNA \cite{kl:JCP:97,kl:PNAS-BJ-PRL}. 
Various patterns of adsorbed counterions,
including those spiraling through DNA major and minor groves, were considered. 
Thus, the effect of
helically structured separation between negative 
and positive charges on each molecule was
rationalized, explaining, in particular, 
a stronger  DNA-DNA attraction in the presence of
counterions preferentially adsorbing into the major groove. 
A number of applications  of the theory \cite{kl:PNAS-BJ-PRL}
proved to be in line with experimental observations, 
and the main properties
of the calculated interaction potential were verified 
by computer simulations \cite{allahyarov:loewen:pre:00}.

Let us draw a plane perpendicular to the parallel axes of the molecules. 
For each molecule draw a
vector joining the axes where the $5'-3'$ strand \cite{sinden} 
hits the plane and call it
`spin'. The angle  between the two spins, 
$\phi$, may be called the angle of mutual azimuthal
orientation of the two molecules. 
A remarkable effect of DNA double strandedness is a peculiar
dependence of the interaction potential on $\phi$.
To a good approximation the  $\phi$-dependent part of the
potential reads \cite{kl:JCP:97} $u(R,\phi)=-A(R)\cos\phi+B(R)\cos^2\phi$,
where $A(R)$ and $B(R)$ are 
positively-definite functions of the interaxial separation $R$, 
with $A(R)$ dominating at large $R$. For this
potential, the optimum angle is
$\hat \phi= \pm\Theta\left(2B(R)-A(R)\right)\cos^{-1}[A(R)/2B(R)]$, 
with the Heaviside step function $\Theta(x)$. 
In other words, there are two symmetrical nonzero values of the
angle at distances smaller than a critical distance at 
which $A(R)=2B(R)$, and zero value of the angle at
large distances. 
The ratio $A(R)/B(R)$ diminishes with 
decreasing $R$ and the absolute value of the angle grows.
The values of the functions $A(R)$ and $B(R)$ depend 
on the parameters of the DNA helical structure
and distribution of adsorbed ions, but typically the 
absolute value of the optimum angle varies
between $0$ and $\pi/2$. 

Thus the problem of statistical properties of 
columnar aggregates of long DNA molecules  can be
mapped on a 2d-problem of XY-spins interacting via such 
an unusual potential. 
Since $A(R)$ and $B(R)$
exponentially decay with $R$, the dominant role is played
by nearest neighbor interactions. While the $\hat \phi=0$-case
is compatible with simple a hexagonal lattice,
the case $\hat \phi \ne 0$ results into
frustrations of positional and orientational order \cite{strey:etal:prl:00}.
Due to the coupling between the
positional and orientational variables in the interaction
(`$R-\phi$ coupling'), one may expect
most peculiar positional and spin structures in the aggregate, 
a feature known as 
mesomorphism of DNA assemblies \cite{podgornik:cocis:84}.

In this work, we analyze the
statistical properties of such assemblies in aqueous solutions. 
We calculate
phase diagrams that depend on the DNA- and salt-concentrations,
and on the 
counterion adsorption pattern. 
To investigate the stability of various phases,   
we carry out lattice-sum calculations for interacting 
DNA molecules and supplement them with the
entropic and cohesive contributions from the ions of the solution. 
The so-obtained variational
Helmholtz free energy is finally minimized among the 
candidate phases and the equilibrium states are
obtained.
Treating DNA molecules as rigid is justified as long as
their contour length does not
exceed the persistence length $L_p = 500\,{\rm \AA}$. The axes of the
molecules remain parallel (to the $z$-axis)
as long as the nearest-neighbor distance
in the aggregate remains below $40\,{\rm \AA}$.
To model the interaction, we envision the molecules as 
long cylinders, carrying helical, continuous
line charges on their surface.
Each DNA-double helix carries the negative  charge  of
phosphates with surface charge density 
$\sigma = 16.8\,{\mu\rm C/cm^2}$ plus a compensating positive charge
coming from the adsorbed counterions. 
Let $0 < \theta < 1$ be the degree of charge compensation,
$f_1$, $f_2$, and $f_3$ be the fractions of condensed counterions
in the major and the minor grooves, and on the two strands,
respectively ($f_1 + f_2 + f_3 = 1$).
The mobile counterions in solution screen the Coulomb
interactions between the helices, causing at large separations an
exponential decay of the latter with the Debye screening length
$\kappa^{-1}$. Solvent screening is accounted for by 
its dielectric constant $\varepsilon$.
For DNA structural parameters we take the B-DNA values: 
pitch $H \approx 34\,{\rm \AA}$
($g = 2\pi/H$) and the hard-core radius $a = 9\,{\rm \AA}$. 
For the pair interaction potential, we take the form \cite{kl:prl:99}
($R > 2 a$):
\begin{eqnarray}
\nonumber
&\frac{u(R,\phi)}{u_0}\;& = \sum_{n=-\infty}^{\infty}
\left[f_1\theta + (-1)^nf_2\theta - (1 - f_3\theta)\cos(n\tilde\phi_s)\right]^2 \\
\nonumber
&\times&\frac{(-1)^n\cos(ng\Delta z)K_0(\kappa_n R) - 
\Omega_{n,n}(\kappa_n R, \kappa_n a)}
         {(\kappa_n/\kappa)^2[K'_n(\kappa_n a)]^2},\\
& &
\label{interaction}
\end{eqnarray}
where $\Delta z$ is a vertical displacement, 
equivalent to a `spin angle' $\phi = g\Delta z$ \cite{foot}.
Here, $u_0 = 8\pi\sigma^2/{\varepsilon\kappa^2}$ 
($\approx 2.9\,k_BT/{\rm \AA}$ at 
physiological ionic strength),
$\tilde\phi_s \approx 0.4\,\pi$ is the azimuthal 
half-width of the minor groove, and $\kappa_n = \sqrt{\kappa^2 + n^2 g^2}$.
$\Omega_{n,m}(x,y)$ is given by
\begin{equation}
\Omega_{n,m}(x,y) = \sum_{j=-\infty}^{\infty}
                    \left[K_{n-j}(x)K_{j-m}(y)\frac{I'_j(y)}{K'_j(y)}\right],
\label{omega}
\end{equation}
with the modified Bessel functions $K_n(x)$ and $I_j(y)$. The primes denote
derivatives.
The precise form of the $\phi$-dependence is affected by the 
distributions $f_i$, $i=1,2,3$ of the condensed counterions \cite{kl:prl:99}. 
Keeping only the $n=0$-term in the sum of Eq.\ (\ref{interaction}) returns
a pair potential depending on $R$ only, as it corresponds to the 
approximation of continuously charged cylinders. 
Truncating the infinite sum of Eq.\ (\ref{interaction}) at $|n|=5$
is sufficient for full convergence of the sum, for all cases studied here.
\begin{figure}
\begin{center}
    \includegraphics[width=6.0cm,height=4.7cm,clip]{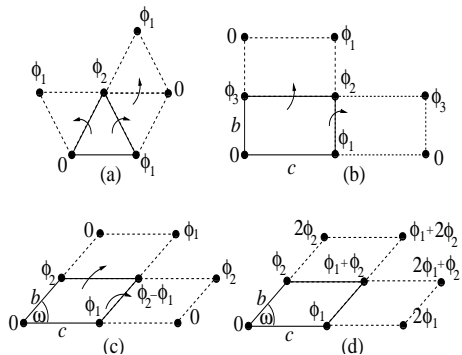}
\caption{A schematic view of generating
         candidate ordered spin
         phases of the system. 
         (a): for the {\textit {HEX}}-lattice; (b): 
              for the {\textit {REC}}- and
                                    {\textit {SQ}}-lattices;
         (c) and (d): fir the {\textit {RHO}}- and {\textit {OBL}}-lattices.}
\label{reflect.fig}
\end{center}
\end{figure}

For all case studies of this work, the pair potential is greater than
$k_BT$, thus we focus on the ground state-analysis of the basic structures
of the assembly.  
To this end, we considered
the five two-dimensional Bravais lattices, i.e., 
the hexagonal ({\textit {HEX}}),
square ({\textit {SQ}}), rectangular ({\textit {REC}}), 
rhombic ({\textit {RHO}}) and oblique ({\textit {OBL}}) lattices.
In order to explore the {\it ordered} spin structures, 
we constructed a certain spin pattern on the elementary plaquette
of every lattice and repeated it along the lattice directions. 
If all site-site interactions in the Hamiltonian are
contained within the elementary geometrical cell 
(plaquette) of the lattice, then the exact ground state can be obtained 
as follows.
The energy of this
plaquette must be minimized with respect to the spin angles,
and then the optimized spin pattern on the plaquette must be
repeated throughout the lattice. 
We have interactions of higher-order-neighbors in our
model but the exponential decay of the $R$-dependent
prefactors guarantees that the nearest-neighbor
interaction dominates. We have kept up to 10 neighboring shells
in the calculations of the lattice sums, that turn out to be
sufficient for convergence.

In Fig.\ \ref{reflect.fig} we show schematically the algorithms employed
for the generation of the ordered spin structures. 
Choosing the orientation of one of the spins
as reference ($\phi = 0$), we are left with two
free orientations per plaquette for the 
{\textit {HEX}}-lattice and three for the
{\textit {REC}}- and {\textit {SQ}}-lattices. 
The lattice is filled by successive mirror-reflections of
the cells across their edges, as shown in Figs.\ \ref{reflect.fig}(a) and (b).
As far as the {\textit {RHO}}- and {\textit {OBL}}-lattices are 
concerned, the 
procedure involving three free spin angles per plaquette does not
generate identical plaquettes upon reflection: 
in these lattice types there is a short and a long diagonal which exchange
their roles upon reflection. We employ two
complementary algorithms for generating ordered magnetic structures on these  
lattices: first, we place spins of orientations $\phi_1$ and $\phi_2$ along
the cell edges and $\phi_1 - \phi_2$ along the long diagonal and use the
successive reflection algorithm. This guarantees that
all pairs of spins across all diagonals will have relative angles
$\phi_1 - \phi_2$, see
Fig.\ \ref{reflect.fig}(c).
Alternatively, we place along the long diagonal a spin with an angle
$\phi_1 + \phi_2$ and subsequently we increase the spin angle along the 
horizontal direction by an amount of $\phi_1$, and along
the oblique direction by an amount of $\phi_2$ for every step. 
We generate thus structures in which all spins along the short diagonals
have an angle $\phi_2 - \phi_1$ and all spins along the long diagonals
an angle $\phi_1 + \phi_2$, as shown in Fig.\ \ref{reflect.fig}(d).  
\begin{figure}
\begin{center}
    \includegraphics[width=6.0cm,height=5.0cm,clip]{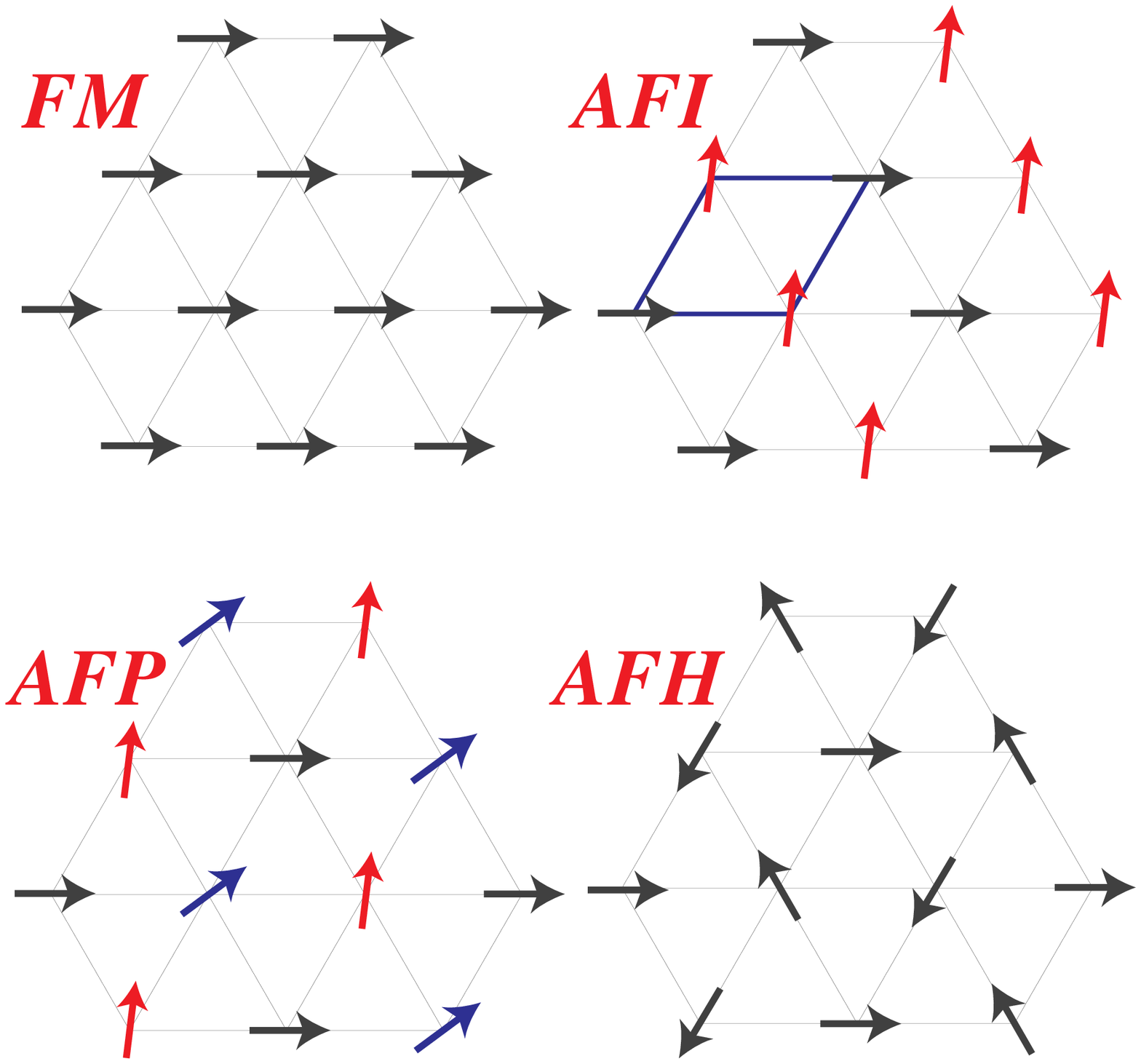}
\caption{The four stable magnetic phases.
         The arrows indicate the relative
         orientations of the helical DNA molecules. The acronyms
         stand for {\it ferromagnetic} ({\textit {FM}}),
         {\it antiferromagnetic Ising} ({\textit {AFI}}),
         {\it antiferromagnetic Potts} ({\textit {AFP}}), and
         {\it antiferromagnetic Heisenberg} ({\textit {AFH}}).}
\label{magnetic.plot}
\end{center}
\end{figure}

The interaction between any two molecules is $L_p\,u(R,\phi)$, where
$L_p = 500{\rm\,\AA}$ is the persistence length and $u(R,\phi)$ is given by 
Eq.\ (\ref{interaction}). It was found that the energy needed 
to destroy 
the translational or orientational order must be more than 
several $k_BT$ at room temperature, hence the lattice-sum calculations
provide the representative thermodynamic states.
The 2d DNA-concentration $\rho$ was varied within 
$0 \leq \rho a^2 \leq 1/(2\sqrt{3})$, the upper limit corresponding to the 
close-packed configuration in a {\textit {HEX}}-lattice. For every density, 
minimizations of the lattice energy with respect to the 
plaquette sets $\{\phi_i\}$, the size ratios $b/c$ 
(for the {\textit {REC}}-lattice)
and/or the geometrical angle $\omega$ ({\textit {RHO}}- and 
{\textit {OBL}}-lattices), Fig.\ \ref{reflect.fig}, were carried out.
This way the 
optimized lattice-sum energy, $U_{\rm X}({\bf \Phi},\rho)$, was obtained,
where 
${\rm X}$ stands for the lattice type and ${\bf \Phi} = (\phi_1,\phi_2,\ldots,\phi_N)$
denotes the configuration of the $N$ spins in
the system. 

To access the full thermodynamics of the DNA solution-salt mixture, we have
to add the contributions to the free energy arising from the counter- and
co-ions, (numbers $N_{\pm}$ and concentrations $c_{\pm}$, respectively.)
The effect of these degrees of freedom
is to add an extensive term to the free energy of the system 
\cite{graf:pre:99}, 
$F_{\rm c} = F_{+}^{0} +  F_{-}^{0} + F_{\rm coh}$, where
$F_{\pm}^{0} = N_{\pm}\,k_BT\,\left[\ln(c_{\pm}\Lambda_{\pm}^3) - 1\right]$
are entropic contributions from the kinetic part of the Hamiltonian
with the thermal de Broglie wavelengths $\Lambda_{\pm}$ 
of the counter- and co-ions, and
\begin{equation}
F_{\rm coh}  =   -\frac{1}{2}\Biggl[
                 \frac{2Na(Ze)^2\kappa}{\varepsilon L_p(1 + \kappa a)} 
  + \frac{k_BT\,V(c_{+}-c_{-})^2}{c_{+}+c_{-}}\Biggr],
\label{cohesive}
\end{equation}
is a cohesive term. In Eq.\ (\ref{cohesive}), $e$ is the
electron charge, 
$Z|e| = 2\pi a L_p\sigma(1-\theta)$ 
is the uncompensated DNA-charge, $c_{+} = Z\rho/L_p + n_s$ and
$c_{-} = n_s$, with
the salt concentration $n_s$. Finally, 
$V$ is the volume of the system and 
$\kappa = \sqrt{4\pi(Z\rho/L_p + 2n_s)e^2/(\varepsilon k_BT)}$ for monovalent
salt ions. 
The Helmholtz free energy is $U_{\rm X} + F_{\rm c}$.
\begin{figure*}
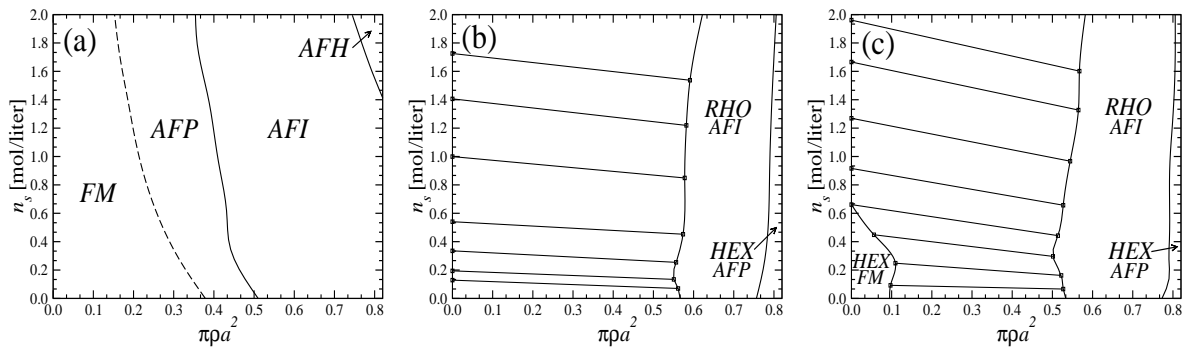

\begin{center}
\begin{minipage}[t]{5.2cm}
    \includegraphics[width=5.0cm,height=4.5cm,clip]{fig3a.eps}
\end{minipage}
\begin{minipage}[t]{5.2cm}
    \includegraphics[width=5.0cm,height=4.5cm,clip]{fig3b.eps}
\end{minipage}
\begin{minipage}[t]{5.2cm}
    \includegraphics[width=5.0cm,height=4.5cm,clip]{fig3c.eps}
\end{minipage}
\caption{Phase diagrams of DNA-salt mixtures obtained by the procedure described
in the text:
(a) $\theta = 0.9$, $f_3=1$. The geometrical lattice is here {\textit {HEX}}. 
(b) $\theta = 0.9$, $f_1=0.3$, $f_2 = 0.7$; 
(c) $\theta = 0.7$, $f_1=0.3$, $f_2 = 0.7$. Dashed lines denote 
second-order magnetic transitions and solid lines first-order ones.
The geometrical transitions between different lattice types
in (b) and (c) are 2nd order; the straight lines 
are tielines between coexisting phases.} 
\label{phdgs.plot}
\end{center}
\end{figure*}

When counterions are condensed {\it on strands}, i.e., $f_1=f_2=0$ and
$f_3=1$, the DNA-DNA interaction is purely repulsive.  The system is
found to crystallize into the {\textit {HEX}} lattice at all
DNA-densities but a large variety of orientational (magnetic)
structures occur, as a result of the frustration of the system.  The
structures are shown in Fig.\ \ref{magnetic.plot} and the phase
diagram of the DNA-salt mixture in Fig.\ \ref{phdgs.plot}(a).  The
phase denoted {\textit {FM}} is a simple ferromagnetic phase, in which
all DNA-molecules have the same azimuthal orientation. The phase
denoted {\textit {AFI}} displays antiferromagnetic-Ising type
ordering, with half DNA-molecules having a given orientation angle if
they lie on one of the sublattices and a different orientation on the
other. The {\textit {AFP}} phase has a three-state antiferromagnetic
Potts \cite{yeomans:book} type of ordering, with 1/3 of the spins
pointing in a reference direction $\phi = 0$, 1/3 in the angle
$\phi_0$ and 1/3 in the angle $2\phi_0$. Note that the angle $\phi_0$
grows with DNA concentration. Finally, the {\textit {AFH}}-phase has
the orientational ordering of the two-dimensional antiferromagnetic
Heisenberg model, with spins residing in the three sublattices of the
hexagonal lattice having mutual orientational angles of $120^{\,\rm
  o}$ to one another. The transition between the {\textit {FM}} and
{\textit {AFP}} phases is second-order but the {\textit {AFP}} $\to$
{\textit {AFI}} and {\textit {AFI}} $\to$ {\textit {AFH}} transitions
are first order.  Referring to Fig.\ \ref{phdgs.plot}(a), we see that
the {\textit {FM}} phase is stable at low DNA-concentrations.  Indeed,
for such average intermolecular separations the optimal orientation
angle between the molecules is zero. The nontrivial phases arise at
higher densities of the aggregates. Similar mesophases were found
recently within the framework of a phenomenological Landau theory
\cite{lorman:prl:01}.

When counterions condense {\it in grooves}, an attraction between the
DNA-molecules arises, since the possibility of having positively
charged parts of one molecule approaching close to negatively
charged parts of the other through an appropriate mutual orientation
opens up. This leads to broad phase coexistence lines between
dense DNA-aggregates and DNA-free solutions. This is demonstrated in 
Fig.\ \ref{phdgs.plot}(b) for the case 
$f_1=0.3$, $f_2=0.7$ and $f_3=0$ for $\theta = 0.9$.
Lowering $\theta$, i.e., increasing the Coulomb repulsion, opens up a 
hexagonal phase at small DNA- and electrolyte-concentrations,
see Fig.\ \ref{phdgs.plot}(c).
In the one-phase region, 
a rhombic phase shows up for 
moderate to high densities and, due to packing constraints, 
a {\textit {HEX}} crystal appears at very high DNA-concentrations. 
A strong {\it qualitative} difference in the
macroscopic behavior of columnar DNA assemblies arises, 
depending on whether the counterions condense on strands or in grooves.
In the former case, all transitions are `magnetic' in nature. 
In the latter, 
DNA-bundling takes place and rhombic lattices are 
stabilized.   

The predictions of the theory ask for experimental verification. 
Such a task is not easy, since the reliable data up
to date refer only to highly concentrated phases
\cite{grimm:physicab:91},
where the number of the basic
assumptions inherent to the form of the pair potential may be questioned 
(the Debye-Bjerrum approximation, independence of
solvent dielectric constant on the aggregate density, 
effects of nonlocal polarizabilty, etc.).  
Such a verification
might become possible with the increase of experimental resolution 
in X-ray diffraction, which could open the
way for the study of less dense aggregates. 
The predicted specific effect of cation adsorption on the 
phase diagram is particularly challenging. Since the
adsorption isotherms and the distributions 
of the adsorbed ions between the minor and major grooves are 
poorly known, one
should concentrate here on the qualitative effects, i.e.,
the (dis)appearance of mesophases triggered  by
the presence of different DNA condensing counterions. 

The authors are thankful to A.\ Cherstvy, S.\ Leikin, A.\ Parsegian, 
and A.\ Esztermann  
for numerous useful discussions and acknowledge financial support
through the DFG, grants  LO 418/6 and KO 139/4.


\begin{thebibliography}{10}

\bibitem{manning:biophys:78} G.\ S.~Manning, Q.\ Rev.\ Biophys.\ {\bf 11}, 179 (1978).

\bibitem{wison:79} R.\ W.~Wison and
V.\ A.~Bloomfield, Biochem. {\bf 18}, 2192 (1979); 
J.\ Widom and R.\ L.~Baldwin, J.\ Mol.\ Biol.\ {\bf 144}, 431 (1980);
P.\ G.~Heath and J.\ M.~Schurr, Macromolecules {\bf 25}, 4149 (1992).

\bibitem{pelta:96} J.\ Pelta, F.\ Livolant,
and J.-L.~Sikorav, J.\ Biol.\ Chem.\ {\bf 271}, 5656 (1996).

\bibitem{rau:etal:pnas:84} D.\ C.~Rau, B.\ Lee, and V.\ A.~Parsegian,  
Proc. Nat. Acad. Sci. USA {\bf 81}, 2621 (1984).

\bibitem{bloomflield:96} V.~A. Bloomfield, Curr. Opin. Struct. Biol. {\bf 6}, 334 (1996).

\bibitem{tamir:etal:93} A.\ H.\ A.~Tajmir-Riahi {\it et al.}, 
                        J.\ Biomol.\ Struc.\ Dyn.\ {\bf 11}, 83 (1993);
                        I.\  Fita {\it et al.}, J.\ Mol.\ Biol.\ {\bf 167}, 157 (1983); 
                        N.\ V.~Hud {\it et al.}, Biochem.\ {\bf 33}, 7528 (1994);
                        X.\ Shui {\it et al.}, Biochem.\ {\bf 37}, 8341 (1998).

\bibitem{podgornik:cocis:84} 
R.\ Podgornik, H.\ H.~Strey, and V.\ A.~Parsegian, 
Curr.\ Opin.\ Colloid Interface Sci.\ {\bf 3}, 534 (1984). 

\bibitem{kl:prl:99} A.\ A.~Kornyshev and S.\ Leikin, Phys.\ Rev.\ Lett.\ 
                    {\bf 82}, 4138 (1999).
                  
\bibitem{kl:JCP:97} A.\ A.~Kornyshev and S.\ Leikin, J. Chem. Phys. {\bf 107}, 3556 (1997);
                    Erratum, {\it ibid.} {\bf 108}, 7035 (1998).

\bibitem{kl:PNAS-BJ-PRL} A.\ A.~Kornyshev and S.\ Leikin, 
                    Proc. Nat. Acad. Sci. USA {\bf 95}, $13\,597$ (1998);
                    Biophys.\ J.\ {\bf 75}, 2513 (1998);
                    Phys. Rev. Lett. {\bf 86}, 3666 (2001).

\bibitem{allahyarov:loewen:pre:00} E.\ Allahyarov and
                                   H.\ L{\"o}wen, Phys.\ Rev.\ E {\bf 62}, 5542 (2000).

\bibitem{sinden} R.\ R.~Sinden, {\it DNA Structure and
                 Function} (Academic, New York, 1994).

\bibitem{strey:etal:prl:00} 
                 H.\ H.~Strey {\it et al.}, Phys.\ Rev.\ Lett.\ {\bf 84}, 
                            3105 (2000).

\bibitem{foot} The `two-cosine approximation', 
               $u(R,\phi) = -A(R)\cos\phi + B(R)\cos^2\phi$,
               follows from Eq.\ (\ref{interaction}) by truncating the sum at
               $|n| = 2$. The truncated form already captures the main
               effect of the interaction, i.e., the frustration of the optimal 
               azimuthal angle.

\bibitem{graf:pre:99} H.\ Graf and H.\ L{\"o}wen, 
                      Phys.\ Rev.\ E {\bf 57}, 5744 (1998);
                      {\it ibid.} {\bf 59}, 1932 (1999).

\bibitem{yeomans:book} J.\ M.~Yeomans, 
                       {\it Statistical Mechanics of Phase Transitions} 
                       (Clarendon, Oxford, 1992).

\bibitem{lorman:prl:01} V.\ Lorman, R.\ Podgornik, and B.\ \v{Z}ek\v{s}, 
                        Phys.\ Rev.\ Lett.\ {\bf 87}, 218101 (2001).

\bibitem{grimm:physicab:91} R.\ Langridge {\it et al.}, 
                            J.\ Mol.\ Biol.\ {\bf 2}, 19 (1960);
                  S.\ D.~Dover, J.\ Mol.\ Biol.\ {\bf 110}, 699 (1977);
                  H.\ Grimm and A.\ Ruprecht, Physica B {\bf 174}, 291 (1991).

\end{thebibliography}
\end{document}